\documentclass[12pt]{article}
\usepackage{etex} 
\usepackage[greek,english]{babel}

\usepackage[dvipsnames]{xcolor}

\definecolor{mygreen}{rgb}{0.35, 0.6, 0.15}
\definecolor{myblue}{rgb}{0.3, 0.3, 1.0}
\definecolor{myred}{rgb}{0.9, 0.2, 0.0}
\definecolor{myorange}{rgb}{0.85, 0.45, 0.0}

\usepackage{enumerate}
\usepackage[all]{xy} 
\usepackage{url}
\usepackage{hyperref}

\hypersetup{
	colorlinks=true,
    linktoc=all,
    linkcolor=blue,  
    }

\renewcommand{\baselinestretch}{1.1}

\let\a=\alpha

\newcommand{\be}{\begin{equation}}
\newcommand{\ee}{\end{equation}}
\def\ba{\begin{align}} 
\def\ea{\end{align}}
\newcommand{\bea}{\begin{eqnarray}}
\newcommand{\eea}{\end{eqnarray}}

\newcommand{\beqs}{\begin{eqnarray}}
\newcommand{\eeqs}{\end{eqnarray}}

\def\zb {{\bar z}}
\def\bc {{\bar c}}
\def\bb {{\bar b}}
\def\baa {{\bar a}}

\def\bz{{\overline z}}
\def\A{{\bf A}}
\def\C{{\bf C}}
\def\P{{\bf P}}

\def\nb{{\bar n}}
\def\tf{{\tilde f}}
\def\tg{{\tilde g}}

\newcommand{\tA}{{\bf{\tilde A}}}
\newcommand{\tB}{{\bf{\tilde  B}}}

\usepackage{mathrsfs}
\usepackage[T1]{fontenc}
\usepackage{mathpazo}
\usepackage{setspace}
\usepackage{amsfonts}
\usepackage{amssymb}
\usepackage{amsmath}
\usepackage{epsfig}
\usepackage{latexsym}
\usepackage{color}
\usepackage{graphicx}
\usepackage[latin1]{inputenc}
\usepackage{slashed}
\usepackage{multirow}
\usepackage{comment}
\usepackage{soul}
\usepackage{hyperref}
\usepackage{cite}

\usepackage[titletoc]{appendix}

\def\hybrid{\topmargin -20pt    \oddsidemargin 0pt
        \headheight 0pt \headsep 0pt
        \textwidth 6.25in       
        \textheight 9.25in       
        \marginparwidth .875in
        \parskip 5pt plus 1pt   \jot = 1.5ex}

\hybrid

\def\baselinestretch{1.2}

\catcode`\@=11

\def\marginnote#1{}
\newcount\hour
\newcount\minute
\newtoks\amorpm
\hour=\time\divide\hour by60
\minute=\time{\multiply\hour by60 \global\advance\minute by-\hour}
\edef\standardtime{{\ifnum\hour<12 \global\amorpm={am}%
        \else\global\amorpm={pm}\advance\hour by-12 \fi
        \ifnum\hour=0 \hour=12 \fi
        \number\hour:\ifnum\minute<10 0\fi\number\minute\the\amorpm}}
\edef\militarytime{\number\hour:\ifnum\minute<10 0\fi\number\minute}

\def\draftlabel#1{{\@bsphack\if@filesw {\let\thepage\relax
   \xdef\@gtempa{\write\@auxout{\string
      \newlabel{#1}{{\@currentlabel}{\thepage}}}}}\@gtempa
   \if@nobreak \ifvmode\nobreak\fi\fi\fi\@esphack}
        \gdef\@eqnlabel{#1}}
\def\@eqnlabel{}
\def\@vacuum{}
\def\draftmarginnote#1{\marginpar{\raggedright\scriptsize\tt#1}}

\def\draft{\oddsidemargin -.5truein
        \def\@oddfoot{\sl preliminary draft \hfil
        \rm\thepage\hfil\sl\today\quad\militarytime}
        \let\@evenfoot\@oddfoot \overfullrule 3pt
        \let\label=\draftlabel
        \let\marginnote=\draftmarginnote
   \def\@eqnnum{(\theequation)\rlap{\kern\marginparsep\tt\@eqnlabel}%
\global\let\@eqnlabel\@vacuum}  }

\def\preprint{\twocolumn\sloppy\flushbottom\parindent 2em
        \leftmargini 2em\leftmarginv .5em\leftmarginvi .5em
        \oddsidemargin -.5in    \evensidemargin -.5in
        \columnsep .4in \footheight 0pt
        \textwidth 10.in        \topmargin  -.4in
        \headheight 12pt \topskip .4in
        \textheight 6.9in \footskip 0pt
        \def\@oddhead{\thepage\hfil\addtocounter{page}{1}\thepage}
        \let\@evenhead\@oddhead \def\@oddfoot{} \def\@evenfoot{} }

\def\numberbysection{\@addtoreset{equation}{section}
        \def\theequation{\thesection.\arabic{equation}}}

\def\underline#1{\relax\ifmmode\@@underline#1\else
        $\@@underline{\hbox{#1}}$\relax\fi}

\def\titlepage{\@restonecolfalse\if@twocolumn\@restonecoltrue\onecolumn
     \else \newpage \fi \thispagestyle{empty}\c@page\z@
        \def\thefootnote{\fnsymbol{footnote}} }

\def\endtitlepage{\if@restonecol\twocolumn \else \newpage \fi
        \def\thefootnote{\arabic{footnote}}
        \setcounter{footnote}{0}}  

\catcode`@=12
\relax

\def\figcap{\section*{Figure Captions\markboth
        {FIGURECAPTIONS}{FIGURECAPTIONS}}\list
        {Figure \arabic{enumi}:\hfill}{\settowidth\labelwidth{Figure
999:}
        \leftmargin\labelwidth
        \advance\leftmargin\labelsep\usecounter{enumi}}}
 \relax
\def\tablecap{\section*{Table Captions\markboth
        {TABLECAPTIONS}{TABLECAPTIONS}}\list
        {Table \arabic{enumi}:\hfill}{\settowidth\labelwidth{Table
999:}
        \leftmargin\labelwidth
        \advance\leftmargin\labelsep\usecounter{enumi}}}
 \relax
\def\reflist{\section*{References\markboth
        {REFLIST}{REFLIST}}\list
        {[\arabic{enumi}]\hfill}{\settowidth\labelwidth{[999]}
        \leftmargin\labelwidth
        \advance\leftmargin\labelsep\usecounter{enumi}}}
 \relax
\makeatletter
\newcounter{pubctr}
\def\publist{\@ifnextchar[{\@publist}{\@@publist}}
\def\@publist[#1]{\list
        {[\arabic{pubctr}]\hfill}{\settowidth\labelwidth{[999]}
        \leftmargin\labelwidth
        \advance\leftmargin\labelsep
        \@nmbrlisttrue\def\@listctr{pubctr}
        \setcounter{pubctr}{#1}\addtocounter{pubctr}{-1}}}
\def\@@publist{\list
        {[\arabic{pubctr}]\hfill}{\settowidth\labelwidth{[999]}
        \leftmargin\labelwidth
        \advance\leftmargin\labelsep
        \@nmbrlisttrue\def\@listctr{pubctr}}}
 \relax
\makeatother
\newskip\humongous \humongous=0pt plus 1000pt minus 1000pt

\newif\ifdtup

\relax

\def\be{\begin{equation}}
\def\ee{\end{equation}}
\def\ba{\begin{eqnarray}}
\def\ea{\end{eqnarray}}

\def\a{\alpha}

\def\B{{\bf B}}

\def\no{\noindent}

\def\IR{\relax{\rm I\kern-.18em R}}

\def \half {\textstyle{1\over 2}}
\def \threehalf {\textstyle{3\over 2}}

\def\IR{\relax{\rm I\kern-.18em R}}
\def\IL{\relax{\rm I\kern-.18em L}}

\def\inv{^{\raise.15ex\hbox{${\scriptscriptstyle -}$}\kern-.05em 1}}

\def\zb {{\bar z}}

\begin{document}

\renewcommand{\theequation}{\thesection.\arabic{equation}}
\csname @addtoreset\endcsname{equation}{section}

\newcommand{\beq}{\begin{equation}}
\newcommand{\eeq}[1]{\label{#1}\end{equation}}
\newcommand{\ber}{\begin{equation}}
\newcommand{\eer}[1]{\label{#1}\end{equation}}
\newcommand{\eqn}[1]{(\ref{#1})}
\begin{titlepage}
\begin{center}


${}$
\vskip .2 in

{\large\bf Two anyons on the sphere: nonlinear states and spectrum}

\vskip 0.4in

{\bf Alexios P. Polychronakos$^{\dagger}$ \ {\small{and}} \, St\'ephane Ouvry$^*$}
\vskip 0.1in

\vskip 0.1in
{\em
${}^\dagger$Physics Department, the City College of NY, New York, NY 10031, USA\\
\vskip -.04 cm and\\
\vskip -.04 cm
The Graduate Center, CUNY, New York, NY 10016, USA
}

\vskip 0.15in

{\em
${}^*$LPTMS, CNRS,  Universit\'e Paris-Sud, Universit\'e Paris-Saclay,\\ \indent 91405 Orsay Cedex, France}

\vskip 0.1in

{\footnotesize \texttt  apolychronakos@ccny.cuny.edu\hskip 0.05cm , \hskip 0.2cm stephane.ouvry@lptms.u-psud.fr }


\vskip .5in
\end{center}

\centerline{\bf Abstract}

\no
We study the 
energy spectrum of two anyons on the sphere in a constant magnetic field. Making use of rotational invariance we reduce
the energy eigenvalue equation to a system of linear differential equations for functions of a single variable, a reduction
analogous to separating center of mass and relative coordinates on the plane. We solve these equations by
a generalization of the Frobenius method and derive numerical results for the energies of non-analytically derivable states.

\vskip .4in
\noindent
\end{titlepage}
\vfill
\eject

\newpage

\tableofcontents

\noindent

\def\baselinestretch{1.2}
\baselineskip 20 pt
\noindent


\setcounter{equation}{0}

\section{Introduction}

Anyons are an extension of the ordinary statistics of identical particles in two dimensions \cite{oldguys}. They appear prominently in fractional quantum Hall systems \cite{FQHE}, but are also related to other systems of particles with nonstandard statistics, and in particular the Calogero model, as recently rigorously demonstrated \cite{cool}. As a result, they
have been the object of intensive study.

In spite of the substantial research effort on various aspects of anyon systems, their analytical treatment remains
essentially an open problem. In particular, the energy spectrum of the $N$ anyon system even in otherwise simple
potentials is largely unknown. A large class of exact energy eigenstates, collectively known as linear or analytic states,
has been derived for anyons on the plane in a general quadratic potential and constant magnetic field \cite{polyanyons,others}.
Still, analytic states constitute a set of measure zero in the full Hilbert space of anyons.
The complete set of anyonic energy eigenstates remains elusive.

The situation is even more challenging in spaces of nontrivial geometry or topology, such as the sphere. In a recent companion
paper \cite{emeis} we studied the problem of anyons on the sphere with a constant magnetic field and identified a set
of analytic energy eigenfunctions. However, these states were a smaller subset of the full spectrum than the
corresponding known states on the plane. What is even more frustrating, the one problem that can be fully solved analytically
on the plane, namely two anyons in a magnetic field (and/or a harmonic potential), is still not fully solvable on the sphere. The non separability of center of mass and relative coordinates on the sphere, unlike the plane, is a technical impediment, but it also appears that
the spherical geometry and topology create additional tension and complications for anyons.

In this paper we study the system of two anyons on the sphere, focusing on the derivation and properties of energy 
eigenstates that admit no obvious analytic solution. The relevance of such a study lies on the fact that this minimal situation
contains the seeds of the difficulties with anyons on the sphere, and their resolution may lead to analogous advances
in the $N$ anyon case. At any rate, the problem of two anyons on the sphere remains highly nontrivial and is of mathematical interest. 

The approach we follow in this work is to reduce the system to its bare bones by fully exploiting angular
momentum conservation. This will allow us to reduce the eigenvalue equations to functions of a single variable. As the
resulting equations still appear not to be analytically integrable, we develop a generalized Frobenius method that provides
singularity conditions and a recursive method to derive the solutions. At that point we turn to numerical calculations and
explore the behavior of nonanalytic states and the dependence of their energies on the anyon statistics parameter.
Potential implications of the approach and extensions of methods to $N$-anyon systems are discussed in the conclusions.

\section{General formulation}

Our analysis will rely heavily on the coordinate system, methodology and notation introduced in \cite{emeis}, restricted
here to the specific case of $N=2$ anyons, and the reader is referred to the above reference for a fully detailed treatment.

We consider two anyons of unit charge and of mass $m=1/2$ and place them on a sphere of unit radius traversed by a
positive magnetic field $B$. We will work
in the so-called ``singular gauge,'' in which the Hamiltonian and all other operators are identical in form to those of 
free particles but the wavefunction is multivalued, acquiring a phase $e^{i \pi \alpha}$ upon exchanging the two anyons
on a counterclockwise path. $\alpha \in [-1,1]$ is the statistics parameter, interpolating between 
bosons for $\alpha=0$ and fermions for $\alpha=1$.

If there are no special points on the sphere (no observable Dirac string singularities) the anyon parameter and magnetic
field must satisfy a quantization condition, which for $N$ anyons reads
\be
2B = M + (N-1) \alpha ~~,~~~~M = {\rm integer}
\label{Baquant}\ee
$M$ being a monopole number. In our case we simply have
\be
2B = \alpha+M
\ee
The above condition is essential for isotropy, so that the angular momentum remain a good symmetry.

In terms of the spherical projective complex coordinates for the anyons
\be
z_a = \tan{\theta_a \over 2} ~e^{i \varphi_a} ~,~~~ a = 1,2
\ee
the Hamiltonian (in the symmetric gauge and a particular ordering) is expressed as 
\be
H = - \sum_{a=1}^2 (1+z_a \zb_a )^2 \left(\partial_a - {B \zb_a \over 1+z_a \zb_a} \right)\left({\overline \partial}_a 
+{B z_a \over 1+z_a \zb_a }\right)
\label{HB}
\ee
with $\partial_a \equiv \partial/\partial z_a$, ${\overline \partial}_a = \partial/\partial \zb_a$. 

We will work, instead, with the asymmetric set of complex coordinates \cite{emeis}
\be
z _a~, ~~ u_a = {\bz_a \over 1+ z_a {\bz_a}} ~~~~{\rm so ~that}~~~~1+z_a \bz_a = {1\over 1-z_a u_a}
\ee
We will also extract a magnetic factor from the wavefunction $\phi$, rewriting it as
\be
\phi = (1-u_1 z_1 )^{B} (1-u_2 z_2 )^{B} \, \psi 
\label{magfac}\ee
Note that we are not extracting any multivalued anyonic factor, like $( z_1 - z_2 )^\alpha$.
Therefore, the new wavefunction $\psi$ remains multivalued and anyonic, like $\phi$. The Hamiltonian in the new
variables and acting on $\psi$ becomes
\be
H = u_1^2 \partial_{u_1}^2 +  u_2^2 \partial_{u_2}^2 +2 (B+1) (u_1 \partial_{u_1} + u_2 \partial_{u_2} )
- \partial_{z_1} \partial_{u_1} - \partial_{z_2} \partial_{u_2}
\label{h2}
\ee
and the angular momentum generators, acting on $\psi$, are
\beqs
J_+ &=& z_1^2 \partial_{z_1} +z_2^2 \partial_{z_2} + \partial_{u_1} + \partial_{u_2} - 2 z_1\,  u_1 \partial_{u_1} 
-2 z_2 \, u_2 \partial_{u_2} - 2 B (z_1 + z_2 )\nonumber\\
J_- &=&  -\partial _{z_1} - \partial_{z_2} \label{JJ}\\
J_3 &=& z_1 \partial_{z_1} + z_2 \partial_{z_2} - u_1 \partial_{u_1} - u_2 \partial_{u_2} - 2 B\nonumber
\eeqs

Since the above Hamiltonian and angular momentum operators are the sum of two single-particle contributions 
it would be trivial to find their eigenstates, if the anyons were not effectively coupled through their nontrivial braiding
properties. As it stands, the energy eigenvalue problem involves four coupled coordinates. Our task
will be to reduce this problem to one variable by taking advantage of the angular momentum symmetry.

\section{Making use of angular momentum}

We will extract anyon energy eigenstates by focusing our attention to eigenstates of the total angular momentum.
For this to work, rotations must remain a good symmetry, so the magnetic field and anyon parameter have to
satisfy the quantization condition $2B = \alpha + M$ stated before.
The minimal magnetic field that we can have is $B=\alpha/2$. (The case where there is no
quantization and there is a special point, usually the south pole, where a Dirac string carrying the flux of
anyons plus magnetic field enters the sphere cannot be treated with the present method.)

The total angular momentum $j$ of the system is necessarily integer: the single-particle angular momentum (for properly quantized
magnetic field) is integer or half-integer, thus the two-particle angular momentum for fermions or bosons is integer.
Since $j$ is quantized, by continuity in $\alpha$ the anyon angular momentum must be integer as well. For brevity,
we will call $j$ "spin" from now on.

\subsection{Identifying spin-$j$ states}

We will look for "bottom" (lowest weight) states $\psi$ of a spin-$j$ multiplet,
with $j$ a non-negative integer, satisfying
\be
J_- \psi = 0 ~,~~~ J_3 \psi = - j \, \psi
\ee
From the form of $J_-$ in (\ref{JJ}) we deduce that $\psi$ is a function of $u_1, u_2$ and $z = z_1 - z_2$ only. From $J_3 =j$
we see that $\psi$ satisfies
\be
z \partial_z \psi - u_1 \partial_{u_1} \psi - u_2 \partial_{u_2} \psi = (2B - j) \psi
\label{j}
\ee
The general solution of this equation is
\be
\psi = f( z u_1 , z u_2 ) \, z^{2B-j}
\label{bottom}
\ee
where $f( \cdot\,, \cdot )$ is an arbitrary function of two variables. Note that the above $\psi$ has automatically the
correct anyon statistics, provided we choose $f$ to be bosonic or fermionic, since $2B-j = \alpha + M-j$.
For $f$ bosonic/fermionic $M-j$ must be even/odd, respectively, in order to have anyon statistics $\alpha$.
However, in the sequel we will not insist on the
previous conditions, therefore considering both ``$\alpha$-even'' and ``$\alpha$-odd'' states, picking up a phase of
$e^{i\pi\alpha}$ or $-e^{i\pi\alpha}$, respectively, under particle exchange.

This form for $\psi$ would also appear to prejudice our wavefunctions to be of the
"analytic" type $\sim z^\alpha \phi$, rather than the "anti-analytic" type $\sim (u_1 - u_2 )^{-\alpha} \phi$,
but this is not so, as it depends on the specific form of the function $f$. For example, choosing
the bosonic function $f = [(z_1 - z_2 )( u_1 - u_2 )]^{n-\alpha} = (z u_1 - z u_2 )^{n-\alpha}$, we obtain
\be
\psi = (z u_1 - z u_2 )^{n-\alpha} ~z^{\alpha +m-j} = (u _1 - u_2 )^{n-\alpha} ~z^{n+m-j}
\ee
and $\psi$ is of the "anti-analytic" type.

If the wavefunctions (\ref{bottom}) were all acceptable Hilbert space states then they would necessarily be
the bottom states of some $j$ representations of rotations. However, this is not necessarily true, as the action of $J_i$ on them
may not close, leading to non-normalizable states. To further determine their
form we must also implement the "closure" condition, valid for a spin-$j$ representation,
\be
J_+^{2j+1} \psi = 0
\label{top}
\ee

We start by examining the sector $j=0$, in which case the "top" condition (\ref{top}) is simply $J_+ \psi = 0$.
Using the form (\ref{JJ}) for $J_+$ and applying it
on $\psi$ as in (\ref{bottom}) with $j=0$ leads to
\be
- x \partial_x f + y \partial_y f = 0 ~,~~~ x = 1-z u_1 ~, ~ y = 1+ z u_2
\label{0xy}
\ee
Expressing the general function $f(x,y)$ as a function of the new variables $s$ and $q$, where
\be
s = x y ~,~~~ q = \sqrt{y \over x} 
\ee
the above equation becomes simply $\partial_q f = 0$, and its general solution is
\be
f = f(s) = f\bigl( (1-z u_1 )(1+z u_2 ) \bigr)
\ee
In terms of the original particle coordinates the variables $x, y, s$ are
\be
x = {1 +  {\bar z}_1 z_2 \over 1 + z_1 {\bar z}_1} ~,~~ y = {1 + z_1 {\bar z}_2 \over 1 + z_2 {\bar z}_2}
~,~~~ s =  {(1 + z_1 {\bar z}_2 )(1 + z_2 {\bar z}_1 ) \over (1 + z_1 {\bar z}_1) (1 + z_2 {\bar z}_2)}
\label{xys}\ee
For later use, we also define the variable $t=1-s$
\be
t = 1-s = {(z_1 - z_2 ) ({\bar z}_1 - {\bar z}_2 ) \over  (1 + z_1 {\bar z}_1) (1 + z_2 {\bar z}_2)}
\label{tt}
\ee 
$s$ and $t$ are both real and range from $0$ to $1$.
Under particle exchange $x \leftrightarrow y$ while $s,t$ are invariant. Also, as the particles
coincide, 
$z_1 \to z_2$, $s \to 1$, $t \to 0$, while as the particles become antipodal, $z_1 \to -1/{\bar z}_2$, $s \to 0$,
$t \to 1$..

To find the solution of $J_+ \psi =0$ for general $j$ we note that the action of $J_+$ will, in general, create a function that also
depends on $Z = z_1 + z_2$. So we evaluate the action of $J_+$ on sums of monomials of the type
\be
\psi_{n,k} = Z^n \, z^{2B-k} f(q,s)
\ee
We find
\beqs
J_+ \, \psi_{n,k} &=& J_+ \left( Z^n \, z^{2B-k} f(q,s) \right) \nonumber\\
&=&\left( {n \over 2}-k\right) Z^{n+1} \, z^{2B-k} f + Z^n \, z^{2B-k+1} \, q \partial_q f + {n \over 2} Z^{n-1} \, z^{2B-k+2} f\nonumber \\
&\equiv& \left( {n \over 2}-k\right) \psi_{n+1,k} + q \partial_q \psi_{n,k-1} + {n \over 2} \psi_{n-1,k-2}
\label{psipsi}
\eeqs
(in the last line $\partial_q$ acts on the $f(q,s)$ part of $\psi_{n,k}$).
The above expression, together with
\be
J_- \, \psi_{n,j} = n\, \psi_{n-1,j} ~~,~~~~ J_3 \, \psi_{n,j} = -j\, \psi_{n,j}
\label{-3}\ee
provide a realization of the $SU(2)$ algebra in the space of $\psi_{n,k}$ with $\psi_{0,j}$ as lowest weight states.
The existence of a top state satisfying $J_+^{2j+1} \psi = 0$ among the linear combinations of $\psi_{n,k}$, however,
is not guaranteed since $J_+$ and $J_-$ are not necessarily Hermitian conjugates in this space, due to the creation of non-normalizable
states. The existence of top states leads to restrictions on $f(q,s)$.

To find the solutions of the top condition (\ref{top}) we note that $s$ does not participate in equations (\ref{psipsi}, \ref{-3}),
so the dependence on $s$ of the solution is arbitrary. Further, $J_+$ preserves the degree of homogeneity in $q$
of $f(q,s)$. So we restrict our attention to monomials in $q$
\be
f(q,s) = q^m f(s)
\ee
The action of $J_+$ on functions $\psi_{n,k} = Z^n \, z^{2B-k} \, q^m f(s)$, deduced from (\ref{psipsi}), is
\be
J_+\, \psi_{n,k} = \left( {n \over 2}-k\right) \psi_{n+1,k} + m\, \psi_{n,k-1} + {n \over 2} \, \psi_{n-1,k-2}
\ee
This is a lattice type action involving discrete steps in the two-dimensional lattice ($n,k$). Without entering into the details
of the derivation, we can show that $\psi_{0,j}$ will be a solution of $J_+^{2j+1} \psi_{0,j} = 0$ if $m$ takes one of the values
\be
m = -j, -j+1 , \dots , j-1 , j
\ee
Since $J_+^{2j+1}$ is a differential operator of order $2j+1$, the above $2j+1$ functions exhaust the
space of solutions. The most general solution for $f(q,s)$ is, then, the linear combination
\be
f (q,s) = \sum_{m=-j}^j q^m f_m (s)
\label{allm}
\ee
with $f_m (s)$ arbitrary functions of $s$. Substituting $q=(y/s)^{1/2}$ in (\ref{allm}) and (\ref{bottom}),
using variables $x$ and $s$ and renaming $s^{m/2} f_m (s)$ simply $f_{-m} (s)$, the
general expression for the spin-$j$ lowest weight state $\psi_{0,j} \equiv \psi_j$ becomes
\beqs
\psi_j
&=&  z^{2B-j} \sum_{m=-j}^j x^{m} f_m (s) \label{genf}\\
&=& 
z^{2B-j} \bigl[ x^j f_{j} (s) + \dots + x f_{1} (s) + f_{0} (s) + x^{-1} f_{-1} (s) + \dots + x^{-j} f_{-j} (s) \bigr]\nonumber
\eeqs

\subsection{Energy eigenstates}

The Hamiltonian (\ref{h2}) acting on a monomial $z^{2B-j} x^m f_m (s)$ gives
\beqs
H \left(z^{2B-j} x^m f_m \right) \hskip -0.4cm&=\hskip -0.3cm& z^{2B-j} \left\{  x^{m+1} (m-j) f'_m \right. \nonumber \\
&+& \hskip -0.3cm x^m \left[ 2s(s-1) f''_m + 2(2B+m+2) s f'_m -2(m+1) f'_m +m(2B+m+1) f_m \right] \nonumber\\
&-& \hskip -0.3cm\left. x^{m-1} (j+m) (m f + s f'_m ) \right\} 
\label{hxm}\eeqs
with $f'_m = df_m (s)/ds$. The action of the Hamiltonian on the state (\ref{genf}) returns a state of the same form, since it commutes
with $J_i$: when it acts on the state $z^{2B-j} x^j f_j (s)$ the term proportional to $x^{j+1}$ in (\ref{hxm}) vanishes, while
on $z^{2B-j} x^{-j} f_{-j} (s)$ the term proportional to
$x^{-j-1}$ vanishes. The energy eigenvalue equation for the state $\psi_j$ in (\ref{genf}) obtains as
\beqs
&-& \hskip -0.25cm (j+m+1) \Bigl( s f_{m+1}' + (m+1) f_{m+1} \Bigr) \nonumber \\
&+& \hskip -0.25cm 2s(s-1) f_m'' + 2(2B+m+2) \, s f_m' -2(m+1) f_m' +m(2B+m+1) f_m \nonumber \\
 &-& \hskip -0.25cm (j-m+1) f_{m-1}' ~~~=~ E f_m
 \label{Hredj}\eeqs
So the problem of finding energy eigenstates of spin $j$ reduces to solving a system of 
$2j+1$ coupled differential equations for $2j+1$ functions of a single variable $s$. This represents a significant reduction of the
original problem which involved a single wavefunction of four variables.

The above can be considered as the analog of the separation of center of mass and relative coordinates on
the plane. The space symmetries of the planar system are the two magnetic translations and the angular momentum,
and are analogous to the three rotations on the sphere. Fixing the center of mass state is akin to fixing the total momentum
and corresponds to reducing the system by the translation symmetry.
This yields discrete oscillator states, since $[p_x , p_y ] = iB\,$ becomes the Heisenberg group, and choosing the oscillator
ground state is analogous to the condition $J_- \psi=0$ on the sphere. Further fixing the angular momentum $L = J_3$ for the relative states is analogous to choosing $J_3 \psi = j \psi$ on the sphere and leads to an equation for a function of the single
relative radial coordinate $r$, analogous to the multiplet $f_m (s)$ on the sphere.

We can take one more step of reduction, using the particle exchange symmetry of the problem. The transformation
$x \leftrightarrow y$, $s \to s$ implies
\be
f_m (s) \to s^{-m} f_{-m} (s)
\ee
under which (\ref{Hredj}) is invariant. So we can impose the condition (with $2B=\alpha + M$)
\be
f_{-m} (s) = \pm (-1)^{j+M} \, s^m f_m (s)
\ee
the $+$ ($-$) sign corresponding to $\alpha$-even ($\alpha$-odd) states. (\ref{Hredj})
does not mix the two kinds of states. For antisymmetric states $f_{-m} (s) = -s^m f_m (s)$
we are left with the $m>0$ equations in (\ref{Hredj}) with $f_0 (s) =0$, while for symmetric states 
$f_{-m} (s) = s^m f_m (s)$ we have again the $m>0$ equations but with nonzero $f_0 (s)$, plus the $m=0$ equation
\be 
-2 (j+1) ( s f_{1}' + f_{1}) 
+2s(s-1) f_0'' + 2(2B+2) \, s f_0' -2 f_0' = E f_0
\ee
So the energy eigenvalue problem further reduces to two decoupled systems of $j$ and $j+1$ equations. 
For $j=0$ only symmetric states exist.

\subsection{Acceptable solutions}

It remains to see which of the solutions for $\psi_j$ are physically acceptable. The main issue is normalizability,
which can be compromised either by singularities in $\psi_j$ or by its behavior as $|z_1 |$ and $|z_2 |$ go to infinity.

From (\ref{xys}) we obtain for $z_1 , z_2 \to \infty$
\be
x \to {z_2 \over z_1} ~,~~~ y \to {z_1 \over z_2} ~,~~~ s \to 1
\ee
So all the factors in $\psi_j$ in (\ref{genf}) except possibly $z^{2B-j}$ remain finite if $f_m (s)$ is regular at $s=1$.
The prefactor $z^{2B-j}$ is normalizable, due to the magnetic factors in (\ref{magfac}), so $\psi_j$ will be normalizable
if $f_m (s)$ is regular.

The functions $f_m (s)$ can develop singularities only at the boundaries of their domain $s=0$ and $s=1$. These can be
either logarithmic or power law, and we have to see which, if any, are acceptable.

To start, logarithmic divergences in the wavefunction, although formally square integrable, are never acceptable
since they correspond to 
additional delta-function terms in the energy equation. To see it, consider the single free particle wavefunction ($B=0$)
\be
\psi =  {1-z \zb \over 1+ z \zb} \ln(z \zb) +2
\ee
We can check that it satisfies the energy eigenvalue equation with $E=2$, and has normalizable logarithmic
divergences at the north and south pole ($z=0$ and $z=\infty$). However, it is not an acceptable energy
eigenfunction since it is essentially the Green's function for two opposite charges at the north and south pole and its Laplacian
produces delta-functions at these points.

A similar remark applies to wavefunctions with power law singularities, even if normalizable. Assume
\be
\psi \sim z^r ~~~{\rm near~} z=0
\ee
For $r<0$ the above is singular at $z=0$ and can be regularized as
\be
\psi_\epsilon = {z^{r+1} \zb \over z \zb + \epsilon}
\ee
with $\epsilon \to 0$ eventually. The action of the Laplacian gives
\be
\partial_z \partial_\zb \psi_\epsilon = z^r \left( {2\epsilon^2 \over (z\zb + \epsilon )^3}+ {(r-1)\epsilon \over
(z\zb + \epsilon )^2} \right) \longrightarrow_{\hskip -0.65cm _{\epsilon \to 0}}  \; \pi r\, z^r \delta^2 (z)
\ee 
The delta-function
potential vanishes only when the wavefunction vanishes at $z=0$ (that is, $r >0$) or for $r=0$.

The above considerations eliminate solutions where $f_m (s)$ has logarithmic singularities, as these would introduce
spurious delta-function potentials at either coincidence points $z_1 = z_2$ ($s=1$) or antipodal points $z_1 = -1/\zb_2$ ($s=0$).
Similarly, power-law singularities at $s=0$ are not acceptable, since they lead to similarly divergent wavefunctions at
antipodal points $z_1 \zb_2 = -1$.

The power law behavior of $f_m (s)$ at coincidence points $s=1$, $z_1 = z_2$ requires a more careful treatment: the
factor $z^{2B-j}$ in the wavefunction will also vanish or diverge, and we need to examine the full behavior of the
wavefunction. Switching to the variable $t=1-s$, assume that the behavior of $f_m (t)$ at $t=0$ is $\sim t^r$.
From (\ref{tt}) this implies $f_m (t) \sim (z\zb)^r$ as $z \to 0$. The full behavior of $\psi_j$ depends on the 
particle exchange symmetry of the solutions:

a) For symmetric solutions the $f_m (t)$ and $f_{-m}$ parts in the wavefunction contribute a behavior at $t=0$
\be
z^{2B-j} (x^m + y^m ) f_m (t) \sim z^{2B-j}\, (z \zb )^r
\ee
which is regular at $z=0$ if $2B-j+2r \ge 0$.

b) For antisymmetric solutions the corresponding contributions are
\be
z^{2B-j} (x^m - y^m ) f_m (t) \sim z^{2B-j} \, z \, (z \zb )^r
\ee
which is regular at $z=0$ if $2B-j+1+2r \ge 0$. Altogether, if
\beqs
&r \ge {j\over 2} - B ~~~~~~ &{\rm (symmetric)} \nonumber\\
&\hskip 0.21cm r \ge {j-1 \over 2} - B ~~~~~ &{\rm (antisymmetric)} \label{rreg}
\eeqs
the wavefunction $\psi_j$ will be regular. This still leaves the possibility that cancellations among different terms
$f_m $ may improve the behavior and lead to regular $\psi_j$. As we shall see, such situations do occur.
Therefore, among all solutions of (\ref{Hredj}) we must choose the ones that are nonlogarithmic, nonsingular
at $s=0$ and behaving as $t^r$ at $t=0$ with $r$ guaranteeing a regular $\psi_j$ at $z=0$. Condition
(\ref{rreg}) is sufficient but not necessary.

\section{Solutions of the eigenstate equations}

Even with all the reductions using spin and particle symmetry, the eigenvalue equations (\ref{Hredj}) remain quite complex.
Their full analysis requires the application of the Frobenius singularity method around
both ends of the variable $s$, that is, around $s=0$ and $s=1$ ($t=0$), for a system of coupled differential equations.
We will review the generalized Frobenius method for a system of equations, adapted to our situation, and then
proceed to apply it.

\subsection{Frobenius method for a system of equations}
\label{Frob}

Consider a system of $N$ coupled second-order linear homogeneous differential equations with a singular point (that
is, a point at which at least one of the coefficients of second-order derivatives vanishes). We examine the behavior of the
solutions of the system around that point, which we will call $x=0$. Denoting by $F$ the column vector of the $N$
functions, we assume that their behavior is meromorphic around $x=0$. So we write $F$ as
\be
F (x) = \sum_{n=0}^\infty x^{n+r} F_n
\label{mero}
\ee
where $F_n$ are $x$-independent column $N$-vectors and $r$ is the lowest power for which at least one of the components
of $F$
is nonzero (so $F_0 \neq 0$ but $F_{-1}=0$). Note that the shift $r$ from integer powers of $x$ must be the same for
all functions, since they are coupled through the differential equations and we assume that their coefficients do not involve
fractional powers of $x$.

Plugging this
form into the system of differential equations will lead to a set of equations for the coefficients. {\bf Assume} that the
system is such that we can bring these equations to the recursive form
\be
{\A}(n+1+r) F_{n+1} = {\B}(n+r) F_n
\label{recur}
\ee
where $\A(n+r+1)$ and $\B(n+r)$ are $N \times N$ matrices that depend on $n$ as well as the value of $r$. (Such a recursive form is not
guaranteed. Our anyon system, however, is of this special type.) The condition $F_{-1}=0$ gives the relation
\be
\A(r) F_0 = 0
\ee
That is, $F_0$ must be a zero eigenvector of $\A(r)$. For this to be possible for nonvanishing $F_0$ we must have
the degeneracy condition
\be
\det \A(r) = 0
\label{det0}
\ee
Since the system of equations is second order, the diagonal elements of $\A(r)$ are quadratic expressions
in $r$. Therefore, the above is a degree $2N$ polynomial equation in $r$ that has, in general, $2N$ solutions. We have
the following possibilities:

{\it i}) All roots are distinct and no two roots are separated by an integer. This is the simplest case: the corresponding $2N$
null eigenvectors $F_0$ for each $r$ generate $2N$ independent solutions of the equation upon iterating the
recursion relation (\ref{recur}).

{\it ii}) There are some degenerate roots, but they have correspondingly degenerate eigenvectors with the same multiplicity
(double root, 2 eigenvectors etc.). This case presents no problem either: We still have $2N$ independent vectors $F_0$
generating the $2N$ independent solutions of the system.

{\it iii}) There are multiple roots with a degeneracy of eigenvectors smaller that their degree. This can happen for
non-Hermitian matrices: the geometric degeneracy (the number of eigenvectors) can be smaller that the
algebraic degeneracy (the multiplicity of roots). This is a "problem" case, as we do not recover enough solutions.

{\it iv}) Some roots are separated by an integer. This is also a "problem" case: assume that two roots are 
$r$ and $r+k$ with $k$ a positive integer. Then the $n=k-1$ recursion equation for the first root will read
\be
\A(r+k) F_k = \B(r+k-1) F_{k-1}
\label{rint}
\ee
Since $r+k$ is a root, $\det \A(r+k) =0$ and this equation generically is inconsistent and gives no solution for $F_k$. 
It will be consistent
only is $\B(r+k-1) F_{k-1}$ happens to be in the range of $\A(r+k)$ (an "accidental" occurrence). Otherwise, it requires
$F_{k-1} = 0$, which means $F_0 = 0$. So the solution corresponding to $r$ does not actually exist. Instead, the
above equation (\ref{rint}) becomes the $n=-1$ equation for the root $r+k$, fixing $F_k$ to be a null vector for
$\A(r+k)$ and reproducing only one solution.  Again, we do not recover enough solutions.

The resolution of both "problem" cases is that the system in these cases develops additional logarithmic
(non-meromorphic) solutions. By putting
\be
F(x) = F_1 (x) + \ln x ~ F_2 (x)
\ee
into the system, with $F_1$ and $F_2$ meromorphic of the form (\ref{mero}) with $r$ the bigger root,
$r+k$, we find that $F_2$ will satisfy the same equation as before and $F_1$ will satisfy a similar one but
involving also $F_2$ that will also have a unique solution.

The general conclusion is that every "missing"
solution from the recursion relations (\ref{recur}) corresponds to an additional logarithmic solution.

\subsection{Application to the two-anyon problem}
\label{Fran}

To apply the Frobenius method to the anyon problem we need to expand the eigenvalue equation (\ref{Hredj}) around its two
singular points, the coincidence point $s=1$ ($t=0$) and the antipodal point $s=0$, and identify the behavior of solutions.
The condition that regular solutions at $s=1$ extend to regular solutions at $s=0$ determines the energy eigenvalues.

a) Antipodal point $s=0$: We expand $f_m (s)$ as a power series around $s=0$
\be f_m (s) = \sum_{n=0}^\infty s^{r+n} \tf_{m,n}
\ee
Equation (\ref{Hredj}) implies the recursion relation for the coefficients $\tf_{m,n}$
\beqs
&-&\hskip -0.25cm 2(\nb+1)(\nb+m+1) \tf_{m,n+1} - (\nb+1)(j-m-1) \tf_{m-1,n+1}  \nonumber\\
&+&\hskip -0.25cm [2 \nb (\nb +m+2B+1)+m(2B+m+1)-E ] \tf_{m,n} \label{recs} \\
&-&\hskip -0.25cm (j+m+1)(\nb+m+1) \tf_{m+1,n} =0  \nonumber
\eeqs
where for brevity from now on we denote
\be
\nb = n+r
\ee
So (\ref{recs}) is indeed of the form (\ref{recur}) with matrices
\beqs
\tA(\nb+1 )_{m,k} &=& (\nb+1)\bigl[2(\nb+m+1) \delta_{m,k} + (j-m+1) \delta_{m-1,k} \bigr] \nonumber \\
\tB(\nb )_{m,k} &=& \bigl[2\nb(\nb+m+2B+1)+m(2B+m+1) - E\bigr]\delta_{m,k} \label{ABt} \\
&&- (j+m+1) (\nb+m+1) \delta_{m+1,k} \nonumber
\eeqs
Putting $n=-1$, or $\nb +1 = r$, the roots $r$ are given by the solutions of the equation
\be
\det \tA(r) =0
\ee
Since $\tA$ is triangular, its eigenvalues are the diagonal elements and we obtain
\be
r = 0 ~~{\rm(}2j+1~{\rm degenerate)}~,~~~~ r=-j, -j+1, \dots , j
\ee
The positive roots can actually be eliminated by using the particle-exchange symmetry.
For symmetric or antisymmetric states the $\tA$ matrix is truncated to the top left $(j+1) \times (j+1)$
or $j \times j$ component, respectively, and only the negative and zero roots survive.

As found in section {\bf \ref{Frob}}, the negative root solutions (and any logarithmic solutions they would produce, due to
the fact that they differ by integers) are not acceptable. For $r=0$ the entire matrix $\tA$ vanishes, so we obtain
$j+1$ (symmetric) or $j$ (antisymmetric) acceptable solutions, all regular and analytic at $s=0$.

b) Coincidence point $t=0$: In terms of the variable $t=1-s$ the eigenvalue equations for $f_m (t)$ are
\beqs
&-& \hskip -0.25cm (j+m+1) \Bigl( (t-1) f_{m+1}' + (m+1) f_{m+1} \Bigr) \nonumber \\
&+& \hskip -0.25cm 2t(t-1) f_m'' + 2(2B+m+2) \, t f_m' -2(2B+1) f_m' +m(2B+m+1) f_m \nonumber \\
 &+& \hskip -0.25cm (j-m+1) f_{m-1}' ~~=~ E f_m
 \label{Ht}\eeqs
 Expanding $f_m (t)$ as a power series around $t=0$
 \be
 f_m (t) = \sum_{n=0}^\infty t^{r+n} f_{m,n}
 \ee
 and plugging in (\ref{Ht}) we obtain the recursion relation for the coefficients $f_{m,n}$
 \beqs
&-&\hskip -0.25cm (\nb+1)\bigl[(j+m+1) f_{m+1,n+1}-2 (2B+\nb+1) f_{m,n+1} + (j-m+1) f_{m-1,n+1} \bigr] \nonumber \\
&+&\hskip -0.25cm [2 \nb (\nb +m+2B+1)+m(2B+m+1)-E ] f_{m,n} \label{rect} \\
&-&\hskip -0.25cm (j+m+1)(\nb+m+1) f_{m+1,n} ~~=~0  \nonumber
\eeqs
This is of the form (\ref{recur}) with the same matrix $\B(\nb ) = \tB (\nb )$ as in case (a) but a new $\A$ matrix
\be
\A(\nb+1) = -(\nb+1)\big[(j+m+1) \delta_{m+1,k} -2 (2B+\nb+1) \delta_{m,k} + (j-m+1) \delta_{m-1,k}\bigr]
\ee
The matrix $\A$ is now tridiagonal, but its determinant can be calculated. $\det \A(r) =0$ yields the roots 
\be
r = 0 ~~{\rm(}2j+1~{\rm degenerate)}~,~~~~ r=-j-2B, -j+1-2B, \dots , j-2B
\label{sroots}\ee
Reduction to symmetric or antisymmetric solutions, in this case, does not eliminate the positive roots as in case (a).
Instead, the second set of roots in (\ref{sroots}) splits as
\beqs
r &=& -j-2B, -j-2B+2, \dots , j-2B ~~~~~~~~~~~~ {\rm (symmetric)}\nonumber \\
r &=& -j-2B+1, -j-2B+3 , \dots , j-2B-1  ~~~{\rm (antisymmetric)}
\eeqs
Again, for $r=0$ the entire matrix $\A$ vanishes and we obtain $2j+1$ solutions. For the nonzero roots we may obtain
either logarithmic solutions, since they differ by integers, or power law solutions if ''accidental'' conditions hold.

Acceptable solutions depend on the value of $B$. We consider the following cases:

i. $2B \ge j$: the condition (\ref{rreg}) guarantees that all $2j+1$ solutions for $r=0$ are acceptable, while all other violate it.
In this case the problem can be solved explicitly: the solution for $f_m (s)$ is analytic on the entire domain. Starting from
$t=0$ we can impose a vanishing condition on the coefficients that truncates it to a polynomial and guarantees that no
nonanalytic behavior develops at $s=0$. That is, we can impose the condition
\be
\B ( n ) F_n = 0
\ee
for some $n$, which ensures that $F_k = 0$ for $k>n$. For this to be possible we must have
\be
\det \B (n) =0
\ee
The matrix $\B(n) = \tB (n)$ in (\ref{ABt}) is triangular, so this gives the energy as
\beqs
E &=& 2n(n+m+2B+1)+m(2B+m+1) \nonumber \\
&=& n(n+2B+1) + (n+m)(n+m+2B+1) \label{EBbig}
\eeqs
This corresponds to two particles placed at Landau levels $n$ and $n+m$ ($m=0,1,\dots,j$). For $2B=\alpha+M$
the particles are anyons with energies as above and their wavefunction $\psi_j$ is of the analytic type.

ii. $2B=M<j$, $M$ integer: this corresponds to bosons or fermions in a magnetic monopole $M$. In this case the problem
is solved in terms of linear combinations of products of single-particle states with a fixed energy and spin. Only a fraction
of the states are guaranteed by (\ref{rreg}) to be acceptable, so cancellations must occur to give overall $2j+1$ regular
acceptable solutions.

iii. $2B=\alpha +M <j$: this is the most nontrivial case. The particles are anyons and at least some of
the states will be antianalytic. States have to be checked explicitly for regularity, since condition (\ref{rreg}) in general
is too restrictive.

Overall we see that the challenging and interesting case is $j>2B=\alpha+M$: it includes analytic states as well as all the
antianalytic states of the spectrum and requires a careful analysis to determine the physical states.

\section{Solutions for specific values of $j$}

The general treatment of the spin-$j$ energy eigenvalue problem is quite complicated. We therefore proceed by dealing with
special cases. Since our primary goal is to uncover antianalytic states and study their properties, we will focus on the cases
$j=0$ and $j=1$, which include analytic states as well as the first emergence of antianalytic states.

\subsection{The case $j=0$}

For $j=0$, (\ref{Hredj}) reduces to the single equation for the wavefunction $\psi_0 = f(s) \, z^{2B}$
\be
2 s (s-1) f'' + 4(B+1)\, s f' - 2 f' = E f
\label{Ej0}
\ee
Solutions of the above equation that lead to regular and normalizable states $\psi_0$ contain all spin-$0$ anyon energy eigenstates.

It is instructive to examine equation (\ref{Ej0}) for the case $B=0$ (free fermions or bosons) or $2B = M$ (fermions or bosons
in a magnetic monopole $M$ field) to make contact with know results and understand which anyon states are recovered.

\subsubsection{The case $j=0$, $B=0$}

In the case of noninteracting free particles we can build 2-particle states as tensor products of single-particle ones.
Each particle can be
in a spin-$j$ state, $j=0,1,\dots$, so the total spin is $j_1 \otimes j_2 =( j_1 + j_2 )\oplus (j_1 + j_2 -1) \oplus \cdots \oplus |j_1 - j_2|$, and to obtain a total spin $j=0$ we must have $j_1 = j_2$. States corresponding to total spin $j=0$ are always bosonic,
consistent with the fact that, for $B=0$, $\psi_0 = f(s)$ is symmetric under particle exchange. Equation (\ref{Ej0}) becomes 
\be
s (s-1) f'' + (2 s -1) f' = {E \over 2} f
\ee
From standard analysis, the above equation in general has two independent solutions at least one of which behaves as
$\sim$ $\ln s$ near $s=0$ (compare to Bessel-J and Bessel-K). Eliminating the logarithmic solution leaves one
analytic solution that admits a power expansion around $s=0$ with Taylor coefficients $\tf_n$ satisfying
\be
(n+1)^2 \tf_{n+1} = [n(n+1) - E/2] tf_n
\ee
with $\tf_n =0$ for $n<0$. If positive power coefficients are allowed to remain nonvanishing for all
$n$ they behave asymptotically as $\tf_n \sim 1/n$ and imply a behavior $f \sim \ln (1-s)$ near $s=1$ (the corresponding 
singular solution near $s=1$). To avoid this singularity the series must truncate (compare to harmonic oscillator), and thus
\be
E = 2 n (n+1)
\ee
for some $n$, making $f(s)$ a polynomial of degree $n$, in accordance with our general analysis. 
This is consistent with the fact that a particle of spin $j=n$ has energy $E=J^2 = n(n+1)$ (in units $2m=R=1$) 
and two particles at this level have twice that energy. 

\subsubsection{The case $j=0$, $B\neq 0$}

The situation for $B>0$ is similar. The equation for the Taylor coefficients of the regular solution around $s=0$ is now
\be
(n+1)^2 \tf_{n+1} = [n(n+2B+1) - E/2] \tf_n
\ee
Again the series must terminate, so
\be
E = 2 n (n+2B+1)
\ee
for some non-negative integer $n$, reproducing the energy of two particles at Landau level $n$. 
Note that for the minimal magnetic
field $2B=\alpha$ we obtain the anyonic energies $2n(n+\alpha +1)$. For $2B=\alpha=1$ (fermions in monopole number $M=1$)
the corresponding states $\psi_0 = z f(s)$ become fermionic and they correspond to two fermions at Landau level $n$ of 
spin $j=n+\half$ and total spin $0$.

For ordinary statistics and even monopole number $M$ the single-particle spin $j=n+M/2$ is integer, so there are no fermionic states with
spin $j=0$, as  antisymmetric states in the $j \otimes j$ space for integer $j$ have odd spins $2j-1$, $2j-3$ etc. 
Likewise, for odd monopole number $j=n+M/2$ is half-integer, and spin-$0$ states cannot be bosonic, as symmetric states in 
the $j \otimes j$ space have odd spins. Thus, spin-$0$ states are bosonic for even $M$ and fermionic for odd $M$.

States for $2B = M + \alpha$  are of the form $z^{M+\alpha} f(s)$ with $f(s)$ a symmetric polynomial in the coordinates,
so they are all of the "analytic" type of exact states uncovered in \cite{emeis}. In fact, for a Landau level $n$ state we have
\be
f(s) = s^n + \dots = (1-z u_1 )^n (1+z u_2 )^n + \dots = u_1^n u_2^n (z_1 - z_2 )^{2n} + \dots
\ee
with ellipses standing for lower order terms, so it corresponds to a linear state in \cite{emeis} 
with $P_+ = u_1^n u_2^n$ ~and~ $k=2n+M$.

\subsection{The case $j=1$}

In the more interesting case $j=1$ the states (\ref{genf}) are of the form 
\be
\psi_1 = z^{2B-1} [ x f_1 (s) + f_0 (s) + y f_{-1} (s) ]
\label{ggg}\ee
with $f_{-1,0,1}$ three arbitrary functions of $s$. We consider states symmetric and antisymmetric under
exchange of $x$ ad $y$, corresponding to $\alpha$-even and $\alpha$-odd states for $M$ even, or vice versa for
$M$ odd, and deal with each case separately.

\subsubsection{Antisymmetric case}

In this case $f(x,y) = - f(y,x)$ and thus $f_0 =0$ and $f_1 = - f_{-1} \equiv g$. 
The full state is, up to a coefficient
\be
\psi_{1,{\rm s}} = z^{2B-1} (x-y) g(s) = -z^{2B} (u_1 + u_2 ) g(s)
\ee
so it is again of the analytic type. The eigenvalue equation becomes
\be
2s(s-1) g'' + 2(2B+3) s g' - 4g' + 2(B+1) g = E g
\label{santi}
\ee
This is of a similar type as the $j=0$ one and its solutions, determined by its singularity structure, consist of a
logarithmic one around $s=0$ that we discard and an analytic one with Taylor coefficients satisfying
\be
(n+1) (n+2) g_{n+1} = [n(n+2B+2)+B+1-E/2] g_n
\ee
Similar requirements of regularity at $s=1$ require that the series terminate, which yields the energy eigenvalue condition
\be
E = 2 [n(n+2B +2) +B+1] = n(n+2B+1) + (n+1)(n+2B+2)
\label{Eanti}
\ee
for some $n$, and $g(s)$ becomes again a polynomial of degree $n$ in $s$.
For the minimal case $2B=\alpha$ this is an $\alpha$-even anyonic state.
It corresponds to putting one particle at Landau level $n$ and one at Landau level $n+1$, as is obvious from the way 
we rewrote the energy (\ref{Eanti}) as a sum. For higher monopole numbers these states are bosonic
for even $M$ and fermionic for odd $M$.

Once again, this solution recovers analytic states that have been found in \cite{emeis}, as it is an overall factor $u_1 + u_2$
times a polynomial with leading term $u_1^n u_2^n (z_1 - z_2 )^{2n+M+\alpha}$, which corresponds to the linear
states in \cite{emeis} with $P_+ = u_1^{n+1} u_2^n + u_2^{n+1} u_1^n$ and\break
$k=2n+M$.

\subsubsection{Symmetric case and nonanalytic states}

We finally come to the truly interesting case of symmetric solutions in $x,y$. In this case we have $f_1 = f_{-1} \equiv g$
and $f_0$ does not have to vanish. The states are
\be
\psi = z^{2B-1} [f_0 + (x+y) g] =z^{2B-1} f + z^{2B} (u_2 - u_1 ) g ~,~~~ f\equiv f_0 + 2 g
\label{sym}
\ee
and the energy eigenvalue condition gives two coupled equations
\beqs
&&s(s-1) f'' + (2B+2) s f' -2 f' + 2B g = \epsilon f \nonumber \\
&&s(s-1) g'' + (2B+3) s g' - g' - \half f' + (B+1) g = \epsilon g
\label{ssys}
\eeqs
where $\epsilon = E/2$ is the energy per particle.

Before embarking in the solution of this system let us examine what kind of anyon states it would produce and
what are the regularity requirements for $f$ and $g$.

An inspection of the state (\ref{sym}) shows that for $2B =\alpha +M >1$ it is of the analytic type. For $2B=\alpha <1$, however,
it has a {\bf lower} power of $z = z_1 - z_2$ than the corresponding power of ${\bar z}_1 - {\bar z}_2$.
It is therefore of the "antianalytic" type, the first of this type we encounter.
Further, in this case $z^{2B-1}$ is singular as $z \to 0$ and $f$ must cancel this singularity.

The eigenvalue equations in the variable $t$ are
 \beqs
&&t(t-1) f'' + (2B+2) t f' -2 B f' + 2B g = \epsilon f \nonumber \\
&&t(t-1) g'' + (2B+3) t g' - (2B+2) g' + \half f' + (B+1) g = \epsilon g
\label{tsys}
\eeqs
The task is to find solutions of the system (\ref{tsys}) that lead to a nonsingular (\ref{sym}) at $t=0$ and remain
nonsingular as $t \to 1$.

We can, in fact, find a simple solution of the above equations:
\be
f= 2g = {1 \over s} ~,~~ \epsilon = -B
\ee
satisfies (\ref{tsys}) (this corresponds to $f_0 = 0$ in (\ref{sym})). Similarly, $g=1/s$ in the antisymmetric case of the
previous section also satisfies equation (\ref{santi}) with $E=2\epsilon = -2B$. Unfortunately they are not acceptable
solutions, since the symmetric one diverges as $z_1 \to z_2$ ($s \to 1$), while the antisymmetric one diverges as 
$z_1 \to -1/{\bar z}_2$ ($s \to 0$). A careful analysis is required to identify acceptable solutions.

Expanding $f$ and $g$ around $t=0$ their coefficients satisfy the coupled equations
\beqs
&&(\nb+1)(\nb+2B) f_{n+1} = [\nb (\nb+2B+1)-\epsilon] f_n + 2B g_n  \\
&&(\nb+1) [-\half f_{n+1} + (\nb+2B+2) g_{n+1} ] = [\nb (\nb+2B+2)+B+1 -\epsilon] g_n\nonumber
\eeqs
where $\nb = n+r$, or in matrix form
\be
\hskip -0.3cm (\nb+1) \hskip -0.1cm \left[\begin{matrix}{\hskip -1cm \nb+2B ~~~~~~~~~0}\\
{\hskip 0.2cm -\half ~~~~~~\nb+2B+2}
\end{matrix} \right]\hskip -0.1cm
\left[ \begin{matrix}f_{n+1}\\
g_{n+1}
\end{matrix} \right] =
 \left[\begin{matrix}{\hskip -0.8cm \nb (\nb+2B+1)-\epsilon ~~~~~~~~~~~~~2B}\\
{\hskip 0.2cm 0~~~~~~~~~~\small{\nb^2 + (2\nb+1)(B+1)-\epsilon}}
\end{matrix} \right]\hskip -0.1cm
\left[ \begin{matrix}f_{n}\\
g_{n}
\end{matrix} \right] 
\label{neart}\ee
This is of the general form (\ref{recur}).
The condition (\ref{det0}) on the matrix $\A(r)$ for the existence of solutions (corresponding to $\nb = r-1$ above) gives
\be
r^2 (r+2B-1)(r+2B+1) = 0
\ee
We recover the double root $r=0$ and the two roots separated by an integer, $r=1-2B$ and $r=-1-2B$, consistent
with the general analysis of section {\bf \ref{Fran}} for $j=1$.

For the double root $r=0$, $\A(r)$ vanishes and we can choose $f_0$ and $g_0$ arbitrarily. We
have two independent solutions, behaving at $t\to 0$ as
\be
f(t) = f_0  + \dots  ~,~~~g(t) = g_0 + \dots
\label{t00}
\ee

For the other roots, the solution corresponding to the larger one $r=1-2B$ requires the condition $\A(r) F_0 =0$
which translates to
\be
-\half f_{0} + 2 g_{0} = 0
\label{f0g0}
\ee
This gives a single solution that behaves as
\be
f(t) = {\hat f}_0 \, t^{1-2B} + \dots ~,~~~ g(t) = {1 \over 4} {\hat f}_0 \, t^{1-2B} + \dots
\label{t1B}
\ee
where we denoted the leading coefficient ${\hat f}_0$ to distinguish it from the one for $r=0$.
The smallest root $r=-1-2B$ will give a solution behaving as $t^{-1-2B}$ near $t=0$ that requires $f_0 =0$. An "accidental" condition holds that makes this solution compatible with and different from the $r=1-2B$ one, so we get another power law solution
\be
f(t) = {\hat g}_0 \, t^{-2B} + \dots ~,~~~
g(t) = {\hat g}_0 \, t^{-1-2B} + \dots
\ee
This exhausts the solutions near $t=0$, all of which turn out to be nonlogarithmic. The full solution behaves as
\beqs
&&f(t) = f_0 + {\hat f}_0 \, t^{1-2B} + {\hat g}_0 t^{-2B} + \dots \nonumber \\
&&g(t) = g_0 + {1\over 4} {\hat f}_0 \, t^{1-2B} + {\hat g}_0 t^{-1-2B} + \dots
\label{solt0}
\eeqs
with $f_0$, ${\hat f}_0$, $g_0$, ${\hat g}_0$ four arbitrary parameters. 

For the singular point $s=0$ we work with the equations (\ref{ssys}) for $f(s)$ and $g(s)$. We obtain for
their coefficients $\tf_n$ and $\tg_n$
\be
\hskip -0.4cm (\nb+1) \left[\begin{matrix}{\hskip -0.1cm \nb+2 ~~~~~~~0}\\
{\hskip 0.1cm \half ~~~~~~\nb+1}
\end{matrix} \right]
\left[ \begin{matrix}\tf_{n+1}\\
\tg_{n+1}
\end{matrix} \right] =
 \left[\begin{matrix}{\hskip -1.9cm \nb (\nb+2B+1)-\epsilon ~~~~~~~~~~~~~~~2B}\\
{\hskip 0.0cm ~~~~~~~~~~0 ~~~~~~~~~~~~~~~\small{(\nb+1)^2 + B (2\nb+1)-\epsilon}}
\end{matrix} \right]
\left[ \begin{matrix}\tf_{n}\\
\tg_{n}
\end{matrix} \right]
\label{nears}\ee
The root equation in this case reads
\be
r^3 (r+1) = 0
\ee
So we have a triple root $r=0$ and a single one $r=-1$, again in accordance with the general analysis of section
{\bf \ref{Fran}}.
The root $r=0$ makes $\A(r)$ vanish and we recover two independent solutions
\be
f(s) = \tf_0 + \dots ~,~~~ g(s) = \tg_0 + \dots
\label{s00}
\ee
The $r=-1$ root produces an "accidental" solution with ${\hat \tf}_{0}=2 {\hat \tg}_{0}$, compatible with (and different
from) the two $r=0$ solutions. So, in general, when $\epsilon \neq B$ we have two regular solutions, one singular one
and one additional logarithmic one (from $r=0$). Substituting
\be
f(s) = f_l (s) \ln s + f_r (s) ~,~~~ g_s = g_l (s) \ln s + g_r (s)
\ee
in (\ref{ssys}) we see that $f_l , g_l$ have themselves to be an $r=0$ solution of (\ref{ssys}) and, moreover, that the
constant term of $f_l$ must vanish, $\tf_{l,0} =0$, while $\tg_{l,0}$ is arbitrary. $f_r $ and $g_r$ are regular at $s=0$
and are fully determined in terms of $\tg_{l,0}$. So the general solution for $f(s), g(s)$ is
\beqs
&&f(s) = \tf_0 + {\hat \tf}_0 \, s^{-1} + \dots \nonumber \\
&&g(s) = \tg_0 + {1\over 2} {\hat \tf}_0 \, s^{-1} + \tg_{l,0} \ln s + \dots
\label{sols0}\eeqs
with $\tf_0 , {\hat \tf}_0 , \tg_0 , \tg_{l,0}$ four arbitrary parameters.

It remains to see which of the above solutions near $t=0$ and $s=0$ are physically acceptable. Near $t=0$ the solution
(\ref{solt0}) implies that the full wavefunction (\ref{sym}) behaves as
\be
\psi_{1,{\rm s}} = f_0 \, z^{2B-1} + {\hat f}_0 \, \zb^{1-2B} + g_0 \, z^{2B} \zb + 2{\hat g}_0 \, z^{-1} \zb^{-2B} + \dots
\ee
We see that acceptable solutions depend on $B$: for $2B >1$ we can choose only $f_0$ and $g_0$ nonzero, while
for $2B<1$ we can choose only ${\hat f}_0$ and $g_0$ nonzero. The term ${\hat f}_0$ corresponds to an antianalytic state.

Near $s=0$ only regular solutions are acceptable, and from
(\ref{sols0}) these can have only $\tf_0$ and $\tg_0$ nonzero. 

Determining the energy eigenvalues now proceeds in the standard way:
The acceptable solutions around each singular point will, in general, produce non-acceptable ones once continued
to the other singular point. To avoid that, we must impose conditions on their constants. 

For $2B>1$ the situation is simple and parallels the previous ones encountered for $j=0$ or $j=1$. Only analytic solutions are
acceptable and we can impose a vanishing condition to terminate the series. Solutions are polynomial with energy
given in (\ref{EBbig}).

For $2B<1$ the solutions are not analytic and there is no vanishing condition. In the next section we develop a method
for determining the solution and apply it numerically to recover the nonanalytic energy levels.

\subsection{A method for finding the energy}

We have not been able to find any analytic way to calculate the eigenvalues. Solving the recursion equations numerically
and fixing the energy such that the regularity condition hold appears to be the only available method at this point.
We highlight the relevant facts below, streamline a possible calculation scheme and present numerical results for the
$j=1$ symmetric case with $2B<1$.

We consider the $t=0$ end and adopt the notation $f_{n+r}, g_{n+r}$ in order to keep track of the dependence of the
expansion coefficents on the root $r$. Taking $2B=\alpha$, for $0<\alpha<1$, acceptable solutions are:
\begin{itemize}
\item For $r=0$, the solution that starts with $f_0=0$, $g_0 \neq 0$, for which $\psi \sim z^{\alpha} {\bar z}$
\item For $r=1-\alpha$, the solution that starts with $f_{1-\alpha} = 4 g_{1-\alpha} \neq 0$, for which 
$\psi \sim {\bar z}^{1-\alpha}$
\end{itemize}
The other two solutions, the one that starts with $f_0 \neq 0$ and the one that starts with $f_{-1-\alpha} \neq 0$, are
not acceptable since they lead to singular $\psi$.

For the above two solutions, the coefficients $f_n , g_n$ and $f_{n+1-\alpha}, g_{n+1-\alpha}$ can be calculated
recursively. By inverting the matrix in (\ref{neart}) we have
\beqs
&&\left[ \begin{matrix}f_{\nb+1}\\
g_{\nb+1}
\end{matrix} \right] = {1 \over (\nb+1)(\nb+\alpha)(\nb+\alpha+2)} \, \times \label{mmm}\\
&& \hskip -0.4cm\left[\begin{matrix}{\hskip -2.4cm (\nb+\alpha+2)[\nb (\nb+\alpha+1)-\epsilon] ~~~~~~~~~~~~~~~~~~~~~\alpha(\nb+\alpha+2)}\\
{\hskip 0.6cm  {\half[\nb (\nb+\alpha+1)-\epsilon]}~~~~~~~~~~~~~~~\small{{\alpha \over 2}+
(\nb +\alpha )[(\nb+1)^2 + \alpha (\nb+\half)-\epsilon]}}
\end{matrix} \right]
\left[ \begin{matrix}f_{\nb}\\
g_{\nb}
\end{matrix} \right]\nonumber
\eeqs
or, more compactly,
\be
\left[ \begin{matrix}f_{\nb+1}\\
g_{\nb+1}
\end{matrix} \right] = \C(\alpha,\epsilon)_\nb \, \left[ \begin{matrix}f_{\nb}\\
g_{\nb}
\end{matrix} \right]
\ee
with $\C(\alpha,\epsilon)_\nb$ the matrix in (\ref{mmm}).
Choosing (arbitrary) appropriate initial conditions $f_0 = 0, g_0=1$ for $r=0$ and $f_{1-\alpha} = 4 g_{1-\alpha} =4$ 
for $r=1-\alpha$
we have
\beqs
\left[ \begin{matrix}f_{n}\\ g_{n} \end{matrix} \right] &=& 
\prod_{k=0}^{n-1} \C(\alpha,\epsilon)_k \, \left[ \begin{matrix}0\\ 1
\end{matrix} \right] \equiv \P(\alpha,\epsilon)_n \, \left[ \begin{matrix}0\\ 1
\end{matrix} \right] \nonumber \\
\left[ \begin{matrix}f_{n+1-\alpha}\\
g_{n+1-\alpha}
\end{matrix} \right] &=& \prod_{k=0}^{n-1} \C(\alpha,\epsilon)_{k+1-\alpha} \, \left[ \begin{matrix}4 \\ 1
\end{matrix} \right] \equiv \P(\alpha,\epsilon)_{n+1-\alpha} \, \left[ \begin{matrix}4\\ 1
\end{matrix} \right]
\label{Solna}
\eeqs
The general solution will be a linear combination of the above solutions with arbitrary coefficients,
\be
c \left[ \begin{matrix}f_{n}\\
g_{n}
\end{matrix} \right] + \bc \left[ \begin{matrix}f_{n+1-\alpha}\\
g_{n+1-\alpha}
\end{matrix} \right]
\label{c1c2}\ee
The coefficients $c$ and $\bc$ must be chosen such that the solution remain nonsingular and nonlogarithmic at $t=1$ ($s=0$).

In general, the coefficients $f_n$ and $g_n$ for large $n$ behave as
\beqs
f_n &=& a_0 + {a_2 \over n^2} + \dots \nonumber \\
g_n &=& {a_0 \over 2} +{a_1 \over n} + {b_2 \over n^2} + \dots
\label{largen}\eeqs
corresponding to behavior near $s=0$ ($t=1$)
\beqs
f(t) &=& \sum_n f_n t^n = {a_0 \over 1-t} + {\rm {regular}} \nonumber \\
g(t) &=& \sum_n g_n t^n = {a_0 \over s(1-t)} - a_1 \ln (1-t) + {\rm {regular}}
\eeqs
So the $a_0$ terms reproduce the singular  solution ${\hat \tf}_0 \, s^{-1}$ in (\ref{sols0}), the $a_1$ term
reproduces the logarithmic solution and the remaining terms the regular part. (The term 
$\sim$ $n^{-1}$ in $f_n$ must be absent since $f(s)$ in (\ref{sols0}) has no leading logarithmic term.)
Each of the solutions
in (\ref{c1c2}) will lead to $f_n, g_n$ in (\ref{largen}) with different coefficients $a_0, a_1 , \dots$ and the full solution
will behave as
\beqs
f_n &=& c a_0 + \bc \baa_0  + {c a_2 + \bc \baa_2 \over n^2} + \dots \nonumber \\
g_n &=& {c a_0+ \bc \baa_0 \over 2} + {c a_1 + \bc  \baa_1 \over n} + {c b_2 + \bc \bb_2 \over n^2} + \dots
\label{gens}
\eeqs
with $a_0, a_1, \baa_0, \baa_1$ deduced from the solutions (\ref{Solna}).

To avoid the singular and logarithmic solutions at $s=0$, the $O(1)$ and $O(n^{-1})$ coefficients must vanish; that is
\be
c a_0 + \bc \baa_0  = c a_1 + \bc  \baa_1 = 0
\ee
For this to be possible with nonzero $c,\bc$ we must have the condition
\be
a_0 \baa_1 - a_1 \baa_0 = 0 ~~~{\rm or} ~~~~R = {a_1 \over a_0} - {\baa_1 \over \baa_0}=0
\label{energycon}\ee
Since $a_0, a_1, \baa_0, \baa_1$ depend on the energy $\epsilon$, the equation $R=0$ is an
eigenvalue condition that determines the energy levels.

To estimate the ratio difference $R$ in (\ref{energycon}) from the solution (\ref{Solna}) we consider the difference of ratios
of Taylor coefficients $R_n$
\beqs
{2 \over n} R_n = {{f_n} \over { g_n}} - {f_{n+1-\alpha}
\over { g_{n+1-\alpha}}} &=& {a_0 + {a_2 \over n^2} + \dots \over {a_0 \over 2} + {a_1 \over n} + {b_2 \over n^2} + \dots} - {\baa_0 + {\baa_2 \over n^2} + \dots \over {\baa_0 \over 2} + {\baa_1 \over n} + {\bb_2 \over n^2} + \dots }
\nonumber \\
&=& {2 \over n} \left[{\baa_1 \over \baa_0} - {a_1 \over a_0} + {A_1 \over n} + {A_2 \over n^2} + \dots \right]
\label{raticus}
\eeqs
So $R_\infty = R$ and the eigenvalue condition (\ref{energycon}) is equivalent to the relation
\be
2\lim_{n\to \infty} R_n = \lim_{n\to \infty} n \left( {{f_n} \over { g_n}} - {f_{n+1-\alpha}
\over { g_{n+1-\alpha}}}\right) =0
\label{ene}
\ee
These coefficients are given in (\ref{Solna}) and are functions of $\alpha$ and $\epsilon$. The roots of the above
equation in $\epsilon$ provide the energy eigenvalues.

The convergence of the ratio $R_n$ can be improved by
considering linear combinations of $N+1$ successive values
\beqs
R_{N,n} &=& {1\over N!} \sum_{k=0}^N {N \choose k} (-1)^k (n-k)^N R_{n-k} \nonumber \\
&=& {\baa_1 \over \baa_0} - {\a_1 \over a_0} + {(-1)^N A_{N+1} \over n^{N+1}} + \dots
\label{RimpN}\eeqs
Calculating the value of $R_{N,n}$ for a reasonable value of $N$ and a large value of $n$ gives an estimate
of $\lim_{n\to\infty} R_n$ with accuracy of order $n^{-(N+1)}$. Alternatively, we can obtain estimates of the
energy levels $E_k (n)$ by using $R_n$ as an estimate for $R$:
\be
R_n = 0 ~~\Rightarrow ~~ E = \{ E_k (n) \}
\ee
Since $R_n \to R$ as $n \to \infty$, $E_k (n) \to E_k$. In general,
\be
E_k (n) = E_k + {E_{k,1} \over n} + {E_{k,2} \over n^2} + \dots
\ee
for some constants $E_{k,1}, E_{k,2},\dots$, and we may use a relation analogous to (\ref{RimpN}) to improve the
convergence of $E_k (n)$:
\beqs
E_k (N,n) &=& {1\over N!} \sum_{k=0}^N {N \choose k} (-1)^k (n-k)^N E_k (n-k) \nonumber \\
&=& E_k + (-1)^N \, {E_{k,N+1} \over n^{N+1}} + \dots
\label{EimpN}\eeqs
Calculating $E_k (N,n)$ for a reasonable value of $N$ and a large value of $n$ gives an estimate of the energy levels $E_k$.
 
\subsection{Numerical results and properties of the solutions}

We used Mathematica to calculate the coefficients and solve equation (\ref{ene}) numerically for specific values of 
$\alpha$, confirming the validity of the method by taking $n$ big enough and checking if the solutions for $\epsilon$ converge. We verified that:
\vskip 0.2cm
\noindent
$\bullet~$ The solutions $E_k (n)$ converge as $n$ increases and reproduce the known energy eigenvalues at $\alpha =0$
and $\alpha =1$, interpolating between these values for $0<\alpha< 1$.\\
$\bullet~$ For $0<\alpha<1$ the energies differ from the "na\"ive" values one would get by imposing a vanishing condition
and making the determinant of the matrix $\B$ in the RHS of (\ref{neart}) or (\ref{nears}) vanish.\\
$\bullet~$ The energies universally decrease as $\alpha$ increases from $\alpha=0$ to $\alpha=1$.

The flow of the first four energy eigenvalues with $\alpha$ is given in figure 1. We recall that the states considered are
$\alpha$-odd (see (\ref{sym})) and so $\alpha=0$ corresponds to fermions and $\alpha=1$ to bosons.
\begin{figure}
\hskip 4.5cm
\hskip -2.3cm\includegraphics[scale=0.8]{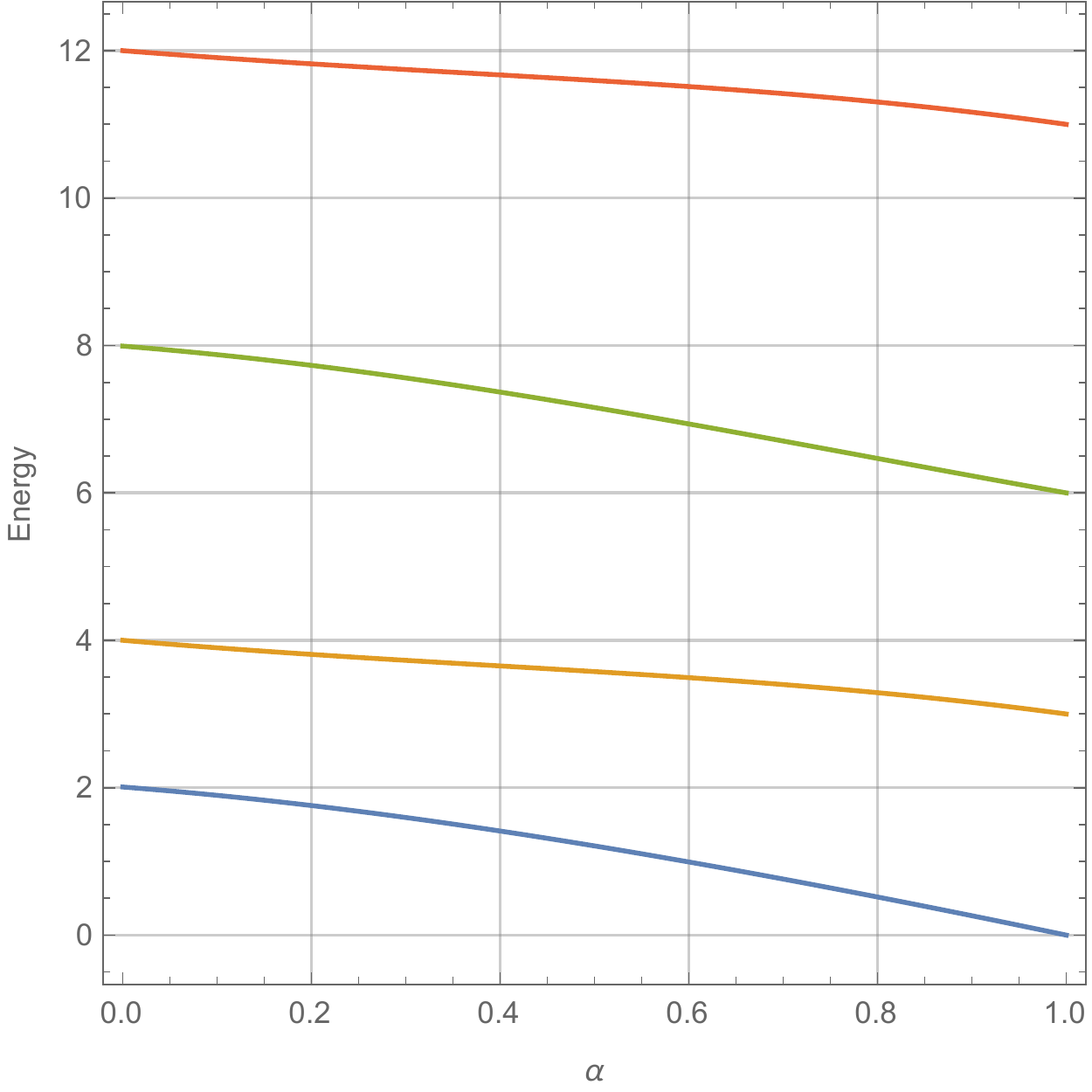}\label{fig1}
\caption{Flow of the lowest four $j=1$ levels from $\alpha=0$ (fermions) to $\alpha=1$ (bosons). They interpolate nonlinearly between Landau levels {\color{myblue}$\mathbf{(0,1) \to (0,0)}$}, {\color{myorange}$\mathbf{(1,1) \to(1,0)}$}, {\color{mygreen} $\mathbf{(1,2) \to(1,1)}$} and {\color{myred}$\mathbf{(2,2) \to(2,1)}$}.}\label{fig1}
\end{figure}
\begin{figure}
\vskip -3.3cm
\hskip -0.5cm
\includegraphics[scale=1.3]{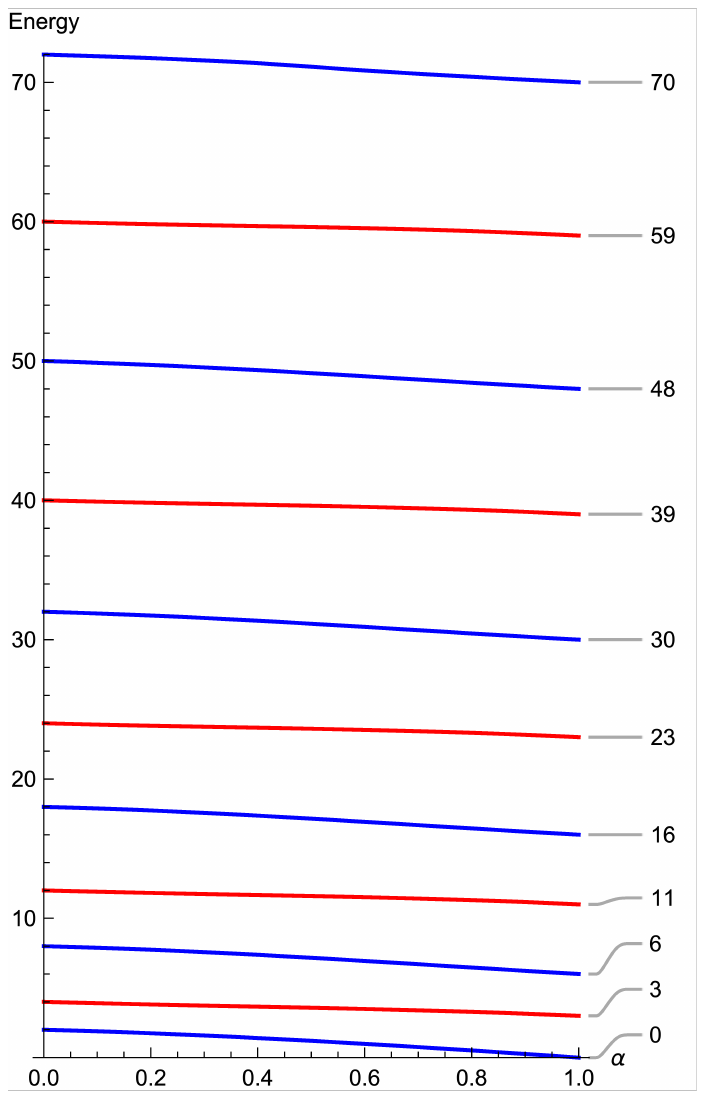}\label{fig2}
\vskip -19.2cm
\caption{Energy level flow for $j=1$ from $\alpha=0$ (fermions) to $\alpha=1$ (bosons). Eigenvalues interpolate between Landau levels
$(n,n+1) \to (n,n)$ ({\color{blue} blue} levels) and $(n+1,n+1) \to(n,n+1)$ ({\color{red} red} levels).}\label{fig1}
\end{figure}
The general interpolation pattern is as in figure 2, between the two anyons being at Landau levels ($n=0,1,2,\dots$):\\
$\bullet~(n,n+1)$ with spins $(n,n+1)$ and energy $n(n+1)+(n+1)(n+2)$ at $\alpha=0$ into $(n,n)$ with spins $(n+\half,n+\half)$ and
energy $2n(n+2)$ at $\alpha=1$, and\\
$\bullet~(n+1,n+1)$ with spins $(n+1,n+1)$ and energy $2(n+1)(n+2)$ at $\alpha=0$ into $(n,n+1)$ with spins $(n+\half,n+\threehalf)$
and energy $n(n+2)+(n+1)(n+3)$ at $\alpha=1$.

Overall we have a validation of the analysis and a numerical approach for solving the problem. We also have a confirmation
that the energy eigenvalues vary nontrivially and nonlinearly with $\alpha$, a hallmark of all nonanalytic states.
\vskip 0.5cm

\section{Conclusions}

The two-anyon problem on the sphere proved surprisingly nontrivial and rich in structure. The angular momentum reduction
of our analysis afforded a substantial simplification, but at the end the nonanalytic solutions could only be accessed through
numerical simulations.

There are various directions in which our results could be extended or improved. A more efficient numerical protocol is
an obvious immediate goal. In particular, it would be interesting to see if solving the original differential equations numerically
would be faster or give the energies more accurately than our asymptotic procedure. However,
numerical accuracy was not central to our considerations.

For theoretical purposes it would be of interest to see if the exact expression of $f_n , g_n$ in (\ref{Solna}) could be calculated explicitly and
their asymptotic behavior determined, thus yielding an explicit algebraic equation for the energy values. Further, the solutions
for
$f(t)$ and $g(t)$ can in general be expressed in terms of hypergeometric functions, so the energy equation could be 
formulated as some
condition on these hypergeometric functions, rather than (\ref{ene}).

More generally, the spectrum for arbitrary higher values of the total spin $j$ could be analyzed. Its study would uncover the full eigenvalue flow in the nonanalytic region
and would presumably lead to a qualitative understanding of the behavior and 'braiding' of the spectrum.

Finally, the reduction method of this paper could be extended to spaces of different symmetry and topology, such as the torus.
The correspondence with the planar model in the limit of infinite radius also remains an intriguing issue.
Planar states would correspond to asymptotically large values of $j$, as the radius $R$ increases. In this context, a perturbative
calculation of the energy levels, considering the spherical Hamiltonian as an $1/R$ perturbation of the planar one, would be
useful and illuminating.

\vskip 0.25cm

\noindent
{\bf{Acknowledgments}}

A.P. acknowledges the hospitality of LPTMS at Universit\'e Paris-Sud (Orsay-Saclay), where this work was initiated. 
A.P.'s research was partially supported by
NSF under grant 1519449 and by an ``Aide Investissements d'Avenir'' LabEx PALM grant (ANR-10-LABX-0039-PALM).

\end{document}